\documentclass[prl,superscriptaddress,twocolumn]{revtex4}%
\usepackage{amsfonts}
\usepackage{amsmath}
\usepackage{amssymb}
\usepackage[dvips]{graphicx}
\usepackage[dvips]{color}
\usepackage{epsfig}
\usepackage{dsfont}%

\newcommand{\half} {\frac{1}{2}}

\newcommand{\sqnbar} {\sqrt{\overline{n}}}
\newcommand{\nbar} {\overline{n}}

\newcommand{\ket}[1]{ {\left| #1 \right\rangle} }
\newcommand{\bra}[1]{ {\left\langle #1 \right|} }
\newcommand{\abs}[1]{ {\left| #1 \right|}}

\begin{document}
\preprint{ }
\title[Collapse and Revival of Entanglement]{Collapse and Revival of Entanglement between Qubits Interacting via a Quantum Bus}
\author{C. E. A. Jarvis}
\email{catherine.jarvis@bristol.ac.uk}
\affiliation{H H Wills Physics Laboratory, University of Bristol, Bristol BS8 1TL, United Kingdom}
\author{D. A. Rodrigues}
\affiliation{School of Physics and Astronomy, University of Nottingham, Nottingham, NG7
2RD, United Kingdom}
\author{B. L. Gy\"{o}rffy}
\affiliation{H H Wills Physics Laboratory, University of Bristol, Bristol BS8 1TL, United Kingdom}
\author{T. P. Spiller}
\affiliation{Hewlett Packard Laboratories, Filton Road, Bristol, BS34 8QZ, United Kingdom}
\author{A. J. Short}
\affiliation{DAMPT, Center of Mathematical Sciences, Wilberforce Road, Cambridge, CB3 0WA,
United Kingdom}
\author{J. F. Annett}
\affiliation{H H Wills Physics Laboratory, University of Bristol, Bristol BS8 1TL, United Kingdom}

\begin{abstract}
We study the dynamics of the Jaynes-Cummings Model for
two level systems (or qubits) interacting with a quantized single
mode electromagnetic cavity (or `quantum bus').  We show that there is a
time in between the collapse and revival of Rabi oscillations  when the state of the qubit sub-system, $\left\vert
\psi\right\rangle_{attractor} $, is largely independent of its initial state. This
generalizes to many qubits the discovery by Gea-Banacloche for the one qubit case. The qubits in such
 `attractor' states are not entangled either with the field or
among themselves, even if they were in the initial state. Subsequently the entanglement between the
qubits revives. Finally, it is argued that the collapse and revival of entanglement and
the persistence of `non-classicality' is a generic feature of multiple qubits interacting via a
`quantum bus'.
\end{abstract}
\maketitle

   At the heart of Quantum Information Science there is the essential quantum
mechanical notion of entanglement. Measures and dynamics of entanglement are
at the centre of much current research as there are many fundamental questions
yet unanswered \cite{Zyczkowski,Eberly07}. Under such circumstances the study of
simple models which feature interesting time evolution of entanglement \cite{konrad, Mintert}, and
yet are tractable, is of particular importance. In this letter we shall
highlight one of these.

   Quantum dynamics of two level systems (also known as qubits), such as spins in
a magnetic field, Rydberg atoms or Cooper Pair Boxes, coupled to a single
mode of an electromagnetic cavity are of considerable interest in connection
with NMR studies of atomic nuclei \cite{Slichter}, Cavity Quantum
Electrodynamics \cite{Berman} and Quantum Computing \cite{NielsenChng} respectively. The simplest model which captures the salient
features of the relevant physics in these fields is the Jaynes-Cummings model
(JCM) \cite{JaynesCum} for the one qubit case and its generalization for multi qubit
systems by Tavis and Cummings \cite{TavisCum}. In this letter we wish
to focus on the interesting dynamics of entanglement between the qubits as
described by these models.

   One of the most interesting and surprising predictions of the JCM is the `collapse
and revival' of Rabi oscillations of the occupation probabilities
for various qubit states as the system evolves, from an initial
state which is a product of a coherent state $\left\vert
\alpha\right\rangle $, for the radiation field, and a generic qubit
state $\left\vert \psi_{N_q}\right\rangle $ \cite{GerryKnight}. It
is central to our present concern that\ such remarkable dynamics
occurs only because both the matter and the cavity field are treated
fully quantum mechanically. Thus, in the language of quantum
computation \cite{NielsenChng}, we may regard the above system as a
collection of qubits interacting via a quantum bus. Indeed our aim
here is to study the `collapse and revival' of entanglement between
non-interacting qubits induced by a quantum bus \cite{Rodrigues}.

For clarity let us recall the multi qubit JCM Hamiltonian, for
which each qubit labelled $i$ can be either in its ground state ${|g_{i}\rangle
}$, with energy $\epsilon_{g,i}$, or its excited state ${|e_{i}\rangle}$, with
energy $\epsilon_{e,i}$. Up to a constant the Hamiltonian may be written in the following conventional form%
\begin{eqnarray}
\label{eq:Nqham}\hat{H}  &  =&\hbar\omega\hat{a}^{\dag}\hat{a}  +\frac{\hbar}{2}\sum_{i=1}^{N_{q}}\Omega_{i}\hat{\sigma}%
_{i}^{z}+\hbar\sum_{i=1}^{N_{q}} \lambda_{i}\left(  \hat{a}\hat{\sigma}%
_{i}^{+}+\hat{a}^{\dag}\hat{\sigma}_{i}^{-}\right) \nonumber\\
\hat{\sigma}^{z}_i&=&\ket{e_i}\bra{e_i}-\ket{g_i}\bra{g_i},\
\hat{\sigma}^{+}_i=\ket{e_i}\bra{g_i},\ \hat{\sigma}^{-}_i=\ket{g_i}\bra{e_i}
\end{eqnarray}
where $\hat{a}^{\dag}$ and $\hat{a}$ are the creation and annihilation
operators of photons with frequency $\omega$, $\hbar\Omega_{i}=\epsilon
_{e,i}-\epsilon_{g,i}$ and $\lambda_{i}$ is the cavity-qubit$_i$ coupling
constant. Here we consider only the case of resonance between
the qubits and the cavity e.g $\omega=\Omega_{i}$ for all $i$ and uniform
coupling $\lambda_{i}=\lambda$.

\begin{figure}[ptb]
\centering
{\epsfig{file=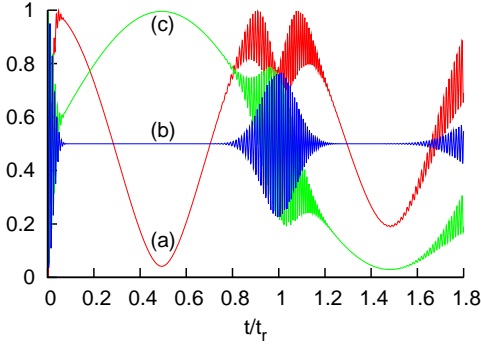, height=5cm} }\caption{(color online) Time evolution for a system with one qubit. (a) the entropy
of the qubit. (b) the probability of being
in the qubit's initial state ${\left|  g \right\rangle } $.  (c) the probability of being in the state ${\left|
\psi\right\rangle } _{attractor}^{+}$. At $t_{r}/2$ the probability of
being in the `attractor' state goes to one while the entropy goes to
zero. The qubit starts in the initial state $\ket{g}$ and the value of $\bar{n}=50$. }%
\label{fig:oneqexp}%
\end{figure}

The celebrated `collapse and revival' can be observed in the one qubit case. We define the initial state:%
\begin{equation}
\ket{\Psi_{1}(0)} =\ket{\psi_1}\ket{\alpha}
\end{equation}
where ${\left|  \alpha\right\rangle } =e^{- {\left| \alpha\right|
}^{2}/2}\sum_{n=0}^{\infty} \frac{\alpha^{n}}{\sqrt{n!}} {\left|  n
\right\rangle }$, $\alpha= \sqnbar e^{-i\theta}$ and
${|\psi_1\rangle}=(C_{g}{|g\rangle}+C_{e}{|e\rangle})$. $\nbar$ is
the average number of photons in the field. The
Rabi oscillations of the
probability that the qubit is in the initial state at first collapse, on
a time scale of $t_{c}\simeq\frac{\sqrt{2}}{\lambda}$, and then revive at
$t_{r}\simeq\frac{2\pi\sqrt{\overline{n}}}{\lambda}$. This is illustrated in
Fig. \ref{fig:oneqexp} for $C_{g}=1$, $C_{e}=0$ by plotting $\sum_{n=0}^{\infty} {\left|  {\left\langle g,n |\Psi_{1}(t)\right\rangle }
\right|  }^{2}$ where $\left\langle g,n\right\vert $ is the state for the
qubit in the ground state with $n$ photons in the cavity.

A second notable feature of this time evolution, discovered by Gea-Banacloche \cite{Ban90},
is that at $\frac{1}{2}t_{r}$, $\left\vert \Psi_{1}(\frac{1}{2}%
t_{r})\right\rangle $ again factorises into a qubit part ${|\psi\rangle
}^+_{attractor}$ and a cavity part ${|\Phi(\frac{1}{2}t_{r})\rangle}$. Moreover,
remarkably, the former is given by%
\begin{equation}\label{eq:1qattractor}
{|\psi\rangle}^{\pm}_{attractor}=\frac{1}{\sqrt{2}}\left(  e^{-i\theta}{|e\rangle
}\pm i{|g\rangle}\right)
\end{equation}
where $\theta$ is the phase of the initial coherent state, for \emph{all} initial
conditions such that $\left\vert C_{g}\right\vert ^{2}+\left\vert
C_{e}\right\vert ^{2}=1$. Note also ${|\psi\rangle}^{-}_{attractor}$ is attained at $t=3t_r/2$. Because of this strikingly non-linear behavior,
following Phoenix and Knight \cite{PheKnight91}, we shall refer to these states, Eq. (\ref{eq:1qattractor}), as
`attractors'. The probability that the qubit is in the state ${|\psi\rangle
}_{attractor}^+$, as depicted by
$\sum_{n=0}^{\infty}\left\vert \left\langle \psi_{attractor}^+,n \vert\Psi_{1}(t)\right\rangle
\right\vert ^{2}$ is also shown in Fig. \ref{fig:oneqexp} together with the von
Neumann entropy $S^Q(t)=-\text{Tr}\left(  \rho^{Q}(t)\ln\rho^{Q}(t)\right)  $
associated with the reduced density matrix $\rho^{Q}(t)=\text{Tr}_F\left(\ket{\Psi_{1}(t)}\bra{\Psi_{1}(t)}\right)$ of the qubit by tracing over the field. Clearly, at $t={\frac{1}{2}t_{r}}$ the entropy
$S^Q(t)$ tends to zero and attains it as $\nbar \rightarrow \infty$, indicating that the radiation field and the qubit are not
entangled \cite{Ban90}.

Prompted by these results we have investigated the $N_{q} >1$ qubit evolution, starting in the state $\ket{\Psi_{N_q}(0)}=\ket{\psi_{N_q}}\ket{\alpha}$, and
found that the spin coherent states \cite{Radcliffe}%
\begin{equation}
\label{eq:attract}{|\psi_{N_q}\rangle}_{attractor}^{\pm}=\frac{1}%
{\sqrt{2^{N_{q}}}}\left(  e^{-i\theta}{|e\rangle}\pm i{|g\rangle}\right)
^{\otimes{N_{q}}}%
\end{equation}
can also be regarded as `attractors' in a similar, dynamical, sense as
outlined above. The only difference is that in the $N_{q}>1$ case
${|\psi_{N_{q}}\rangle }_{attractor}^{\pm}$ will occur only for a
restricted range of initial conditions which we shall call basin of attraction. At the attractor time there is no entanglement between the qubits and the radiation field, furthermore because ${|\psi_{N_{q}}\rangle }_{attractor}^{\pm}$ is a product of one qubit states, the qubits are not entangled with each other.
Below we explore the implications of this observation for the
dynamics of entanglement between the qubits.

\begin{figure}[ptb]
\centering
\epsfig{file=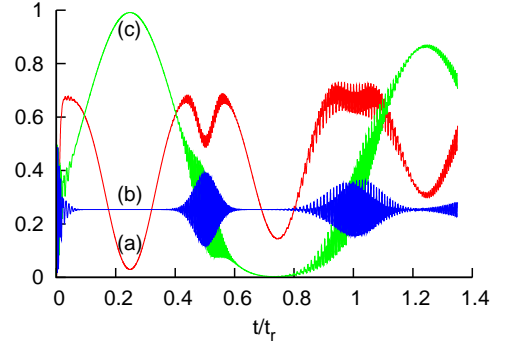,height=5cm}\caption{(color online) Time evolution for a system with two qubits. (a) the entropy of the
qubits. (b) the probability of
the two qubit state ${\left|  gg \right\rangle } $. (c) the probability of being in the two qubit `attractor'
state ${\left|  \psi_2\right\rangle } _{attractor}^+$ when the initial phase of
the radiation field is $\theta=0$. The two qubit `attractor' state is reached at $t_{r}/4$.
The initial state of the qubits is
$\frac{1}{\sqrt{2}}( {\left|  ee \right\rangle } + {\left|  gg \right\rangle }
)$ and the
value of $\nbar=50$.}%
\label{fig:twoqubitatt}%
\end{figure}

From the above point of view the simplest case of interest is that
of two qubits e.g. $N_{q}=2$. In this case the time evolution
described by $\left\vert \Psi_{2}(t)\right\rangle $ is readily found
\cite{Chumakov95}. For the most general, normalized, initial state
\begin{equation}
\left\vert \psi_2\right\rangle =C_{ee}\left\vert ee\right\rangle +C_{eg}%
\left\vert eg\right\rangle +C_{ge}\left\vert ge\right\rangle +C_{gg}\left\vert
gg\right\rangle
\end{equation}
of the qubit sector the exact analytical solution will be given
elsewhere \cite{Jarvis08}. Here we consider only the sector determined by
the restrictions: $a=e^{i\theta}C_{ee}=e^{-i\theta}C_{gg}$ and $\sqrt{\frac{1}{2}-\abs{a}^2}=C_{eg}=C_{ge}$. As will be
illustrated presently these define the `basin of attraction' for the `attractor'
${|\psi_2\rangle}_{attractor}^{+}$. Namely, for any values of $a$ satisfying $0\leq\left\vert a\right\vert \leq1/\sqrt{2}$ in%
\begin{equation}
\label{eq:initattract}{\left|  \psi_2\right\rangle } =a\left(
e^{-i\theta} {\left|  ee \right\rangle } +e^{i\theta} {\left|  gg
\right\rangle } \right)  +\sqrt{\frac{1}{2}- {\left|  a \right|
}^{2}}\left(  {\left|  eg \right\rangle } + {\left|  ge
\right\rangle } \right)
\end{equation}
the probability that the two qubits are in the state
${|\psi_2\rangle }_{attractor}^{+}$ (given by $P_{2\,
attractor}(t)=\left\langle
\psi_{2\,attractor} ^{+}\right\vert \rho^{Q}(t)|\psi_{2\,attractor}^{+}%
\rangle$), will reach 1 at some time $t^{\ast}$. An example of such
behavior (for $\theta=0$) is shown in Fig. \ref{fig:twoqubitatt}. To
highlight the similarity with the analogous phenomena in the one
qubit case (Fig. \ref{fig:oneqexp}) we also show the entropy
$S^Q(t)$. This is calculated from $\rho^Q$, the two qubit density
matrix reduced with respect to the cavity field coordinate, which
describes a mixed state for most times $t$. Notably, at
$t^{\ast}=\frac{1}{4}t_{r}$, where $P_{2\,attractor}(t)=1,$ the
entropy $S^Q(t)$ tends to zero in the large $\nbar$ limit,
indicating that the system of two qubits is not entangled with the
field.

\begin{figure}[ptb]
\centering{\epsfig{file=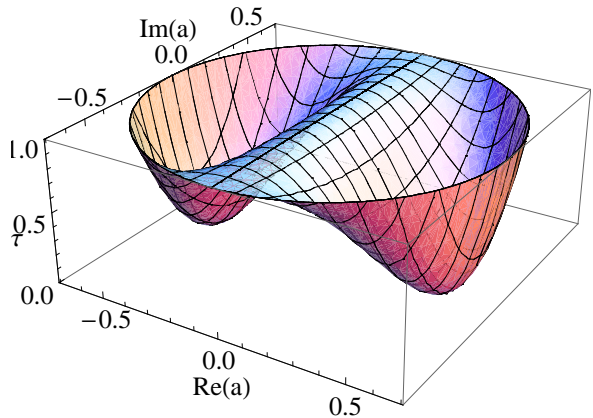,height=6cm} }\caption{(color online) The value of the
tangle for the states in the basin of attraction for different values of
$a$. We notice that there are only two points where the
tangle is zero, $a=\pm\half$.}%
\label{fig:concur}%
\end{figure}

\begin{figure}[ptb]
\centering{\epsfig{file=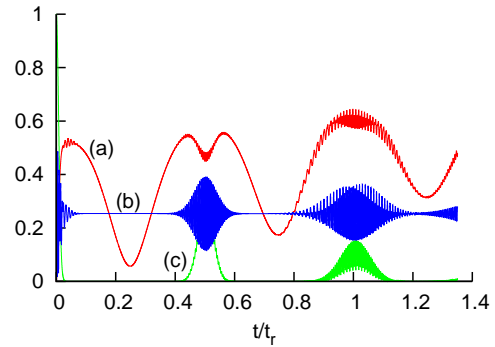,height=5cm} }\caption{(color online) The qubit system
started in the maximally entangled state $({\left\vert ee\right\rangle
}+{\left\vert gg\right\rangle })/\sqrt{2}$ and $\nbar=50$. (a) the entropy of the qubit system. (b) the
probability of being in the state ${\left\vert gg\right\rangle }$. (c) the mixed state tangle of the qubit
system.}%
\label{fig:reventanglement}%
\end{figure}

The interesting new feature of the two qubit case as opposed to the
one qubit case is that the former is in general host to entanglement
between qubits and this provides an opportunity to study the
dynamics of such entanglement. For example, whilst almost all of the
initial states in the `basin of attraction' given in Eq.
(\ref{eq:initattract}) describe entangled qubits they all evolve
into ${|\psi_2\rangle}_{attractor}^{+}$ at $t=\frac{1}{4}t_{r}$
where they are not entangled. We plot the pure state tangle of the
initial condition defined as $\tau=4\left\vert
C_{ee}C_{gg}-C_{eg}C_{ge}\right\vert ^{2} $ \cite{Wooters98} as a
function of $a$ in Fig. \ref{fig:concur}. Note that although there
are only two points where $\tau=0$, all values of entanglement,
including $\tau=1$ meaning maximal entanglement, are present in the
`basin of attraction'. Thus we are observing the time evolution of a
generic amount of entanglement. To throw further light on the
matter we show, in Fig. \ref{fig:reventanglement}, the time
evolution of the mixed state tangle calculated from $\rho^{Q}$ for
the maximally entangled initial state
$|\psi_2\rangle=\frac{1}{\sqrt{2}}\left(  \left\vert ee\right\rangle
+\left\vert gg\right\rangle \right)  $ \cite{Munro}.
Evidently, just as the occupation of the initial qubit states collapses and revives, so does the entanglement.
This phenomenon was first noted by Rodrigues \emph{et al} in a similar
context \cite{Rodrigues}.

Surprisingly, $\tau$ remains near zero for
long periods between revivals. Thus, we are dealing with a phenomenon which
was dubbed the `death of entanglement' by Yu and Eberley \cite{Yu04} and is
the center of much current interest \cite{Eberly07}.  Qing \emph{et al} \cite{Qing} have
found a similar collapse and revival for the same model we have studied
here but for very different initial conditions. What makes our results
even more surprising is that the phenomenon occurs for a well defined range of initial
conditions, namely the `basis of attraction' for all the $N_q$ qubit `attractor' states, and the defining features of these can
be generalized to an arbitrary number of qubits interacting with the same
quantum bus. In fact, using the large $\overline{n}$ expansion of Meunier
\emph{et al} \cite{Meunier} we have found a `basis of attraction' for all the `attractor'
states in Eq. (\ref{eq:attract}) given by%
\begin{eqnarray}
\label{eq:Nqbasin}{\left|  \psi_{N_{q}}\right\rangle }   &  =&\sum_{k=0}%
^{N_{q}} \frac{A(N_q,a) e^{-i (\frac{N_q}{2}-k)
\theta}\sqrt{N_{q}!}}{\sqrt{k!(N_{q}-k)!}} {\left|
N_{q},\frac{N_{q}}{2}-k \right\rangle }\nonumber\\
A(N_q,a)  &  =&\left\{
\begin{array}
[c]{cc}%
a & \text{if $k$ is even}\\
& \\
\sqrt{\frac{1}{2^{N_{q}-1}}- {\left|  a \right|  }^{2}} &
\text{if $k$ is odd}%
\end{array}
\right.
\end{eqnarray}
where $0\leq \abs{a}\leq\frac{1}{\sqrt{2^{N_{q}-1}}}$ and the
states $\left\vert N_{q} ,m\right\rangle $ are the fully symmetrized
$N_{q}$ qubit states. $m$ is the difference between the number of
qubits in the excited state $N_{e}$ and those in the ground states
$N_{g}$.

As noted before, the `attractor' states are manifestly not entangled, but
the states in the basin of attraction are. Although there is no
unique measure of entanglement for $N_{q}>2$ qubits it is reasonable to
assume that an arbitrary $N_{q}$ qubit state in its `basin of the attraction'
is generically entangled. Thus the collapse and revival of entanglement should
be expected to be a generic feature of the $N_{q}$ qubit JCM \footnote{Indeed, we have investigated the 3-qubit case numerically, and observed the revival of a state with GHZ-type pure 3-qubit entanglement \cite{Jarvis08}}.

\begin{figure}[ptb]
\centering{\epsfig{file=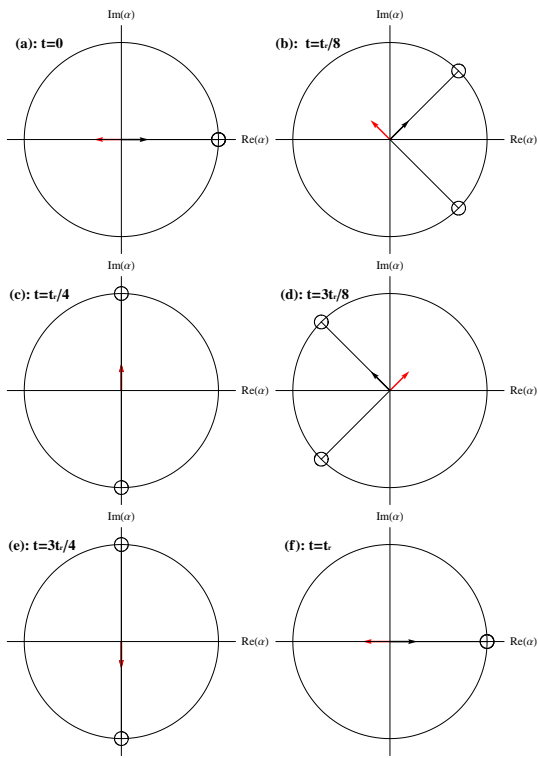,height=10cm} }\caption{(color online) Phase space
sketches of the $Q$ function at six different times when the qubits start in the `basin of attraction'. (a) the time $t=0$, where the
cavity is in a coherent state which is shown by a circle of
uncertainty in phase space. (b) a time a little after $t=0$. (c) the time
$t=t_{r}/4$. (d) just before the time $t=t_{r}/2$. (e) the time $3t_{r}/4$.
(f) the time $t=t_{r}$, when both the circles have returned to their original
position. The qubit dipole states are represented as arrows. The single arrow at $t=t_r/4$ corresponds to the
spin coherent attractor state.}%
\label{fig:2qdipole}%
\end{figure}

Interestingly, unlike in the `two qubit, two cavity' model studied by
Y\"{o}na\c{c}, Yu and Eberly \cite{Yonac}, in the above calculations $\tau$ decays
smoothly to zero with no discontinuities in the gradient, but does not actually go to zero before it revives.
That is to say there is no `sudden death of entanglement' \cite{Eberly07,Qing} and
hence there is no need for `rebirth' \cite{Lopez}. In fact at $t=\frac{1}{4}t_{r}$, when the qubit subsystem is in the `attractor'
state, the entanglement is encoded in the radiation field. At this time both the qubit-resonator entanglement and qubit-qubit entanglement vanish. To investigate the form this
encoding takes we present in Fig. \ref{fig:2qdipole} the $Q$ function
$Q(\alpha,t)=\left\langle \alpha\left\vert \rho^{F}(t)\right\vert
\alpha\right\rangle$, where $\rho^{F}(t)$ is the reduced density matrix for
the radiation field at various times. Note that whilst at $t=0$ and $t_{r}$
there is only one circle which represents a coherent state $\left\vert
\alpha\right\rangle $, at other times there are two circles. At the interesting time $t=\frac{1}{4}t_{r}$, when the radiation field is disentangled from the qubits, there are
two macroscopically different circles on opposite sides of phase space so the state of the cavity is a superposition of the
two coherent states, $\ket{\alpha}$ and $\ket{-\alpha}$.

Such `Schr\"{o}dinger cat' states have been studied by several
authors with various perspectives \cite{Gisin,Ban91,Armour}. Both
`Schr\"{o}dinger cat' states and entangled states may be regarded as
particularly `non-classical' \cite{GerryKnight}, while coherent
states of the field and product states of the qubits are regarded as
more classical states. As a consequence, this fact prompts the
following observation: the entanglement present in the qubit part of
the system at $t=0$ is encoded in the state of the radiation field,
$\ket{\Phi(t)}$, at $t=\frac{1}{4}t_{r}$, which becomes highly
`non-classical'. This is demonstrated by a comparison of $\left\vert
\psi_2\right\rangle $ in Eq. (\ref{eq:initattract}), depicting the
basin of attraction, and the analytic result:%
\begin{eqnarray}
\label{eq:radfield}
\ket{\Phi(\frac{1}{4}t_{r})}&=&e^{i\theta}\left[e^{i\pi \nbar/2}\left(a-\sqrt{\half-\abs{a}^2}\right)\ket{\alpha}\right.\nonumber\\
&+&\left. e^{-i\pi \nbar/2}\left(a+\sqrt{\half-\abs{a}^2}\right)\ket{-\alpha}
\right].
\end{eqnarray}
For example, if the initial qubit state is not entangled, ($\tau=0$), namely $a=\pm \frac{1}%
{2}$, $\ket{\Phi(\frac{1}{4}%
t_{r})}\propto\left\vert \mp \alpha\right\rangle $ then the field state is a more `classical' coherent state. However for $a=\frac{e^{i\phi}}{\sqrt{2}}$, where $\phi$ is an arbitrary phase, or $a=i r$, where $r$ is a real number, the qubits are maximally entangled  ($\tau=1$) and the field
is in the `non-classical' `Schr\"{o}dinger cat' state $\ket{\Phi(\frac{1}%
{4}t_{r})}\propto(\left\vert \alpha\right\rangle +\left\vert
-\alpha\right\rangle )$ characterized by a Wigner function which takes
negative values near the origin. In short
the `non-classicality' which was in the qubit subsystem at $t=0$ is conserved
at  $t=\frac{1}{4}t_r$ when it is in the field subsystem. Remarkably, this also
implies a new strategy for producing `Schr\"{o}dinger cat' states. We shall
elaborate on this interesting possibility in a future publication \cite{Jarvis08}.

Evidently,  whilst we have described the time evolution of entanglement in detail only in the 2-qubit limit
the existence of the `attractor' states, with a
finite basin of attraction, for an arbitrary number of qubits implies that an
oscillatory flow of `non-classicality' between the qubits and the quantum bus,
the cavity, is a generic feature of dynamics described by the multi-qubit
JCM. As the properties of this model are relatively readily accessible,
either analytically or numerically, further study of the `collapse and
revival' of multipartite entanglement dynamics outlined above is clearly
called for. In particular the effect of decoherence on the
\emph{persistence} of non-classicality discovered above remains an open question.

\smallskip
\acknowledgments

The work of C.E.A.J. was supported by UK HP/EPSRC case studentship,
and D.A.R. was supported by EPSRC-GB grant no EP/D066417/1. A.J.S. acknowledges support from a Royal Society University Research Fellowship and the EC QAP project. We thank the ESF network AQDJJ for partial support.

\bibliography{CollapseandRevivalofEntanglement}

\end{document}